\documentclass[12pt]{iopart}

\usepackage{graphicx,times,iopams}
\usepackage[english]{babel}
\usepackage{amssymb}
\usepackage{mathrsfs}
\usepackage{hyperref}
\usepackage{subfigure}
\usepackage[utf8]{inputenc}
\usepackage{cite}
\usepackage{geometry}
\hypersetup{
colorlinks=true,
linkcolor=blue,
citecolor=blue,
filecolor=magenta,
urlcolor=cyan
}

\begin{document}

\title{Strong-coupling effects in dissipatively coupled optomechanical systems}

\author{Talitha Weiss, Christoph Bruder and Andreas Nunnenkamp}

\address{Department of Physics, University of Basel, Klingelbergstrasse 82, CH-4056 Basel, Switzerland}

\begin{abstract}
In this paper we study cavity optomechanical systems in which the position of a mechanical oscillator modulates both the resonance frequency (dispersive coupling) and the linewidth (dissipative coupling) of a cavity mode. Using a quantum noise approach we calculate the optical damping and the optically-induced frequency shift. We find that dissipatively coupled systems feature two parameter regions providing amplification and two parameter regions providing cooling. To investigate the strong-coupling regime, we solve the linearized equations of motion exactly and calculate the mechanical and optical spectra. In addition to signatures of normal-mode splitting that are similar to the case of purely dispersive coupling, the spectra contain a striking feature that we trace back to the Fano line shape of the force spectrum. Finally, we show that purely dissipative coupling can lead to optomechanically-induced transparency which will provide an experimentally convenient way to observe normal-mode splitting.
\end{abstract}

\maketitle

\section{Introduction}

Cavity optomechanical systems have a wide range of possible applications in precision measurement, quantum information, and fundamental tests of quantum mechanics \cite{kv2008, mg2009,aghk2010,m2012,akm2013}. In most optomechanical setups the coupling between the optical and mechanical degrees of freedom arises due to a displacement-dependent cavity frequency (\textit{dispersive coupling}). Driving such systems at a frequency that is red-detuned from the cavity resonance can lead to cooling. This has been theoretically analyzed in Refs.~\cite{mccg2007, w-rnzk2007}, and the quantum ground-state has now been reached in several experiments \cite{t2011, c2011}. In the strong-coupling regime the optical and mechanical degrees of freedom hybridize. Normal-mode splitting has been predicted \cite{dw-rk2008}, and it was subsequently observed \cite{ghva2009} in the optical output spectrum. If the system is probed with an additional probe field, the existence of the two normal modes can lead to destructive interference and a narrow transparency window at the cavity frequency \cite{ah2010}. For dispersive coupling this has been demonstrated experimentally \cite{wrdgask2010,t2011NMS}.

Recently, a different kind of optomechanics has been proposed \cite{egc2009}: a displacement-dependent cavity linewidth leads to a \textit{dissipative coupling} between the mechanical and the optical degrees of freedom. Experimental realizations of this idea have been proposed in the microwave domain for superconducting resonators \cite{egc2009} and in the optical domain for a Michelson-Sagnac interferometer containing a moving membrane \cite{xsh2011}. The ratio between dispersive and dissipative coupling is determined by the position of the membrane. A first experiment that demonstrated a dissipative coupling has been carried out with a microdisk resonator coupled to a nanomechanical waveguide \cite{lpt2009} and, recently, first measurements in an interferometer setup have also been reported \cite{t2012}. It has been pointed out early on that dissipative coupling enables ground-state cooling outside the resolved-sideband limit and has potential applications in quantum-limited position measurements \cite{egc2009}. Moreover, squeezing of the mechanical state \cite{ha2010} and normal-mode splitting in response to a weak probe field \cite{ha2010NMS} have been discussed in the context of the experimental setup of Ref.~\cite{lpt2009}. However, up to date, many properties of dissipatively coupled systems remain unknown.

In this paper we study the general case of a cavity optomechanical system with both dispersive and dissipative coupling. After introducing our model in Section \ref{sec:model}, we examine its mechanical and optical spectra in Section \ref{sec:mechanical} and \ref{sec:optical}, respectively. For weak coupling we employ a quantum noise approach and calculate optically-induced damping and frequency shift of the mechanical oscillator, known as backaction damping and optical spring effect. In contrast to dispersive coupling, we find that dissipatively coupled systems feature two parameter regions of amplification and two parameter regions of cooling. We then present the exact solution to the linearized equations of motion for the general case of dispersive as well as dissipative coupling. If the drive is red-detuned from the cavity resonance by the mechanical frequency, we find signatures of normal-mode splitting in the mechanical and the optical spectra. In the case of dissipative coupling a second feature appears which can be traced back to the Fano line shape in the force spectrum. In Section \ref{sec:OMIT} we discuss optomechanically-induced transparency and find that this interference phenomenon can be observed for purely dissipative coupling. This could be useful to measure the normal-mode splitting.

\section{Model} \label{sec:model}

We consider an optomechanical system that consists of a mechanical oscillator with resonance frequency $\omega_m$ and a cavity mode with resonance frequency $\omega_c$. Dispersive coupling corresponds to a shift of the cavity resonance frequency due to the motion of the mechanical oscillator; dissipative coupling leads to a shift of the cavity damping rate due to the mechanical motion. The Hamiltonian of a dispersively and dissipatively coupled system is given by \cite{egc2009} ($\hbar=1$)
\begin{equation}\label{eqHamiltonian} \fl
\hat{\mathscr{H}} = \omega_c\hat{a}^\dagger\hat{a} + \omega_m\hat{b}^\dagger\hat{b} -\left[ \tilde{A}\kappa\hat{a}^\dagger\hat{a}+i\sqrt{\frac{\kappa}{2\pi\rho}}\frac{\tilde{B}}{2}\sum\limits_{q}^{}\left(\hat{a}^\dagger\hat{b}_q-\hat{b}_q^\dagger\hat{a}\right)\right]\left(\hat{b}^\dagger+\hat{b}\right)+\hat{H}_\kappa+\hat{H}_\gamma .
\end{equation}
The first term describes the cavity mode, where $\hat{a}^\dagger$ ($\hat{a}$) are bosonic creation (annihilation) operators. The second term describes the mechanical oscillator, where $\hat{b}^\dagger$ ($\hat{b}$) are bosonic creation (annihilation) operators. The cavity has a linewidth $\kappa$, and the mechanical oscillator is damped at a rate $\gamma$. The damping due to the optical and mechanical bath is described by $\hat{H}_\kappa$ and $\hat{H}_\gamma$, respectively. The third term describes the optomechanical interaction with dimensionless coupling strengths $\tilde{A}$ (dispersive) and $\tilde{B}$ (dissipative), which are defined as derivatives with respect to the oscillator position $x$ and given by $\tilde{A}\kappa = -\frac{d\omega_c(x)}{dx}x_0$ and $\tilde{B}\kappa =\frac{d\kappa(x)}{dx}x_0$, respectively. Here, $x_0=(2m\omega_m)^{-1/2}$ denotes the size of the zero-point fluctuations and $m$ the mass of the mechanical oscillator; $\hat{b}_q^\dagger$ ($\hat{b}_q$) are bosonic creation (annihilation) operators describing the optical bath coupled to the cavity and $\rho$ denotes the density of states of the optical bath, treated as a constant for the relevant frequencies. $\tilde{B}=0$ corresponds to the well-investigated case of purely dispersive coupling and $\tilde{A}=0$ to the case of purely dissipative coupling.

To derive the Heisenberg equations of motion, we adapt the input-output formalism \cite{wm} to dissipative coupling. This leads to the following expression
\begin{equation}\label{eqIOsum}
\sqrt{\frac{\kappa}{2\pi\rho}}\sum\limits_{q}^{}\hat{b}_q=\sqrt{\kappa}\hat{a}_\mathrm{in}+\frac{\kappa}{2}\hat{a}+\frac{\kappa}{2}\frac{\tilde{B}}{2}\left(\hat{b}+\hat{b}^\dagger\right)\hat{a},
\end{equation}
where $\hat{a}_\mathrm{in}$ is the optical input mode \cite{cdgms2010}. The input-output relation is given by \cite{xsh2011}
\begin{equation}\label{eqIO}
\hat{a}_\mathrm{in}-\hat{a}_\mathrm{out}=-\sqrt{\kappa}\hat{a}-\frac{\sqrt{\kappa}\tilde{B}}{2x_0}\left(\hat{b}+\hat{b}^\dagger\right)\hat{a}.
\end{equation}
Note that the last term only contributes for nonzero dissipative coupling and introduces an explicit dependence on the mechanical displacement as well as a nonlinearity into the input-output relation. 

Using $\hat{a}=(\bar{a}+\hat{d})e^{-i\omega_d t}$, $\hat{b}=\bar{b}+\hat{c}$, $\hat{a}_\mathrm{in}=(\bar{a}_\mathrm{in}+\hat{\xi}_\mathrm{in})e^{-i\omega_d t}$, and Eq.~(\ref{eqIOsum}), we obtain the linearized equations of motion in a frame rotating at the drive frequency $\omega_d$
\begin{equation}\label{eqLinEOMc} \fl
\dot{\hat{c}} = -\left(i\omega_m+\frac{\gamma}{2}\right)\hat{c} -\sqrt{\gamma}\hat{\eta}+i\tilde{A}\kappa\left(\bar{a}^*\hat{d}+\bar{a}\hat{d}^\dagger\right)-\frac{\tilde{B}}{2}\sqrt{\kappa}\left(\bar{a}^*\hat{\xi}_\mathrm{in}-\bar{a}\hat{\xi}_\mathrm{in}^\dagger\right)-i\frac{\tilde{B}}{2}\left(\Omega^*\hat{d}+\Omega\hat{d}^\dagger\right),
\end{equation}
\begin{equation}\label{eqLinEOMd}\fl
\dot{\hat{d}} = i\left(\Delta+\tilde{A}\kappa\frac{\bar{x}}{x_0}\right)\hat{d}-\frac{\kappa}{2}\left(1+\tilde{B}\frac{\bar{x}}{x_0}\right)\hat{d}-\sqrt{\kappa}\left(1+\frac{\tilde{B}}{2}\frac{\bar{x}}{x_0}\right)\hat{\xi}_\mathrm{in}+\left(i\tilde{A}\kappa\bar{a}-\frac{\kappa}{2}\tilde{B}\bar{a}-i\Omega\frac{\tilde{B}}{2}\right)\frac{\hat{x}}{x_0}.
\end{equation}
In this expression, $\Delta=\omega_d-\omega_c$ is the detuning between drive and cavity frequency, $\Omega=-i\sqrt{\kappa}\bar{a}_\mathrm{in}$ is the strength of the coherent laser drive, and $\hat{x}=x_0(\hat{c}^\dagger+\hat{c})$ is the displacement of the mechanical oscillator relative to its steady-state position $\bar{x}$.
The thermal noise influencing the mechanical oscillator is described by the noise operators $\hat{\eta}$ and $\hat{\eta}^\dagger$. The bath coupled to the mechanical oscillator is assumed to be Markovian and at a temperature $T$ associated with an equilibrium phonon number $n_\mathrm{th}=[\mathrm{exp}(\omega_m/k_B T)-1]^{-1}$, i.e.~$\langle\hat{\eta}^\dagger(\omega)\hat{\eta}(\omega')\rangle =  2\pi\delta(\omega+\omega')n_\mathrm{th}$ and $\langle\hat{\eta}(\omega)\hat{\eta}^\dagger(\omega')\rangle = 2\pi\delta(\omega+\omega')(n_\mathrm{th}+1)$ where $k_B$ denotes Boltzmann's constant. The operators $\hat{\xi}_\mathrm{in}$ and $\hat{\xi}_\mathrm{in}^\dagger$ describe the noise induced by the optical bath which is assumed to be vacuum noise, i.e.~$\langle\hat{\xi}_\mathrm{in}(\omega)\hat{\xi}_\mathrm{in}^\dagger(\omega')\rangle = 2\pi\delta(\omega+\omega')$.

In Eq.~(\ref{eqLinEOMd}) dissipative coupling $\tilde{B}$ leads to a change in the damping rate $\kappa$, whereas dispersive coupling $\tilde{A}$ leads to a change in the detuning $\Delta$. These shifts can be determined from the steady-state solutions of the classical equations of motion 
\begin{eqnarray}
0=\dot{\bar{b}}&=-\left(i\omega_m+\frac{\gamma}{2}\right)\bar{b}+i\tilde{A}\kappa\left|\bar{a}\right|^2-i\frac{\tilde{B}}{2}\left(\Omega\bar{a}^*+\Omega^*\bar{a}\right),\label{eqClassEOMb}\\
0=\dot{\bar{a}}&=i\left(\Delta+\tilde{A}\kappa\frac{\bar{x}}{x_0}\right)\bar{a}-\left(1+\frac{\tilde{B}}{2}\frac{\bar{x}}{x_0}\right)i\Omega-\left(1+\tilde{B}\frac{\bar{x}}{x_0}\right)\frac{\kappa}{2}\bar{a}\label{eqClassEOMa},
\end{eqnarray}
where $\bar{x}=x_0(\bar{b}+\bar{b}^*)$. These equations give rise to a static bistability, even if purely dissipative coupling, i.e.~$\tilde{A} = 0$, is considered. This will be discussed elsewhere.

\begin{figure}
\begin{center}
\includegraphics[width=0.9\columnwidth]{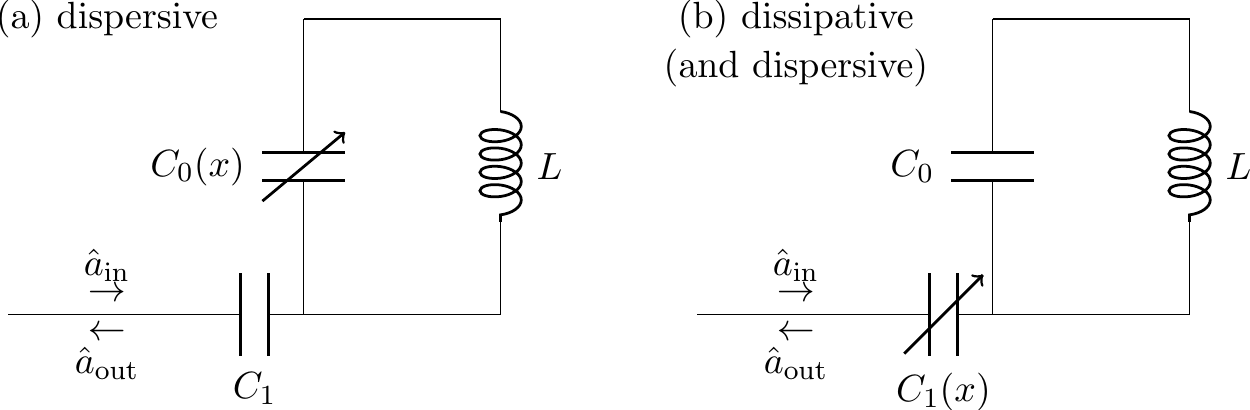}
\caption{(a) Electromechanical implementation of a dispersively coupled system ($\tilde{B}=0$). The resonance frequency of the $LC$ resonator depends on the displacement-dependent capacitance $C_0(x)$. The static output capacitor $C_1$ determines the coupling strength between resonator and feed line.
(b) Electromechanical implementation of a dissipatively coupled system \cite{egc2009}. Compared to (a) only the role of the capacitors is interchanged. This leads to a displacement-dependent coupling of the circuit to the feed line. As the total capacitance and thus the resonance frequency is displacement-dependent, the dispersive coupling is nonzero also in this case (i.e.~$\tilde{A}\neq 0$ and $\tilde{B}\neq 0$).}
\label{picCircuits}
\end{center}
\end{figure}

Figure \ref{picCircuits} shows electromechanical implementations of a dispersively coupled system and a system that includes both types of coupling \cite{egc2009}. In the case of purely dispersive coupling the interaction of the optical bath and drive with the mechanical oscillator is mediated by the cavity. Dissipative coupling also leads to a cavity-mediated influence on the mechanics, but in contrast to dispersive coupling it is proportional to the drive strength $\Omega$ instead of being proportional to the intra-cavity amplitude $\bar{a}$ (cf.~Eq.~(\ref{eqLinEOMc})). In addition, for dissipative coupling the mechanical displacement enters directly in the coupling to the optical bath, see Fig.~\ref{picCircuits} (b), and thus the mechanical oscillator is also directly influenced by the optical bath. This will lead to several new features in dissipatively coupled systems.

\section{Mechanical spectrum} \label{sec:mechanical}
In the following we investigate the properties of the coupled system by discussing some of its fluctuation spectra. These spectra have the form $S_{kq}(\omega)=\int dt \langle\hat{k}^\dagger(t)\hat{q}(0)\rangle e^{i\omega t}=\int d\omega'\langle\hat{k}^\dagger(\omega)\hat{q}(\omega')\rangle/2\pi$ and, in our case, require the solutions of Eqs.~(\ref{eqLinEOMc}) and (\ref{eqLinEOMd}).

In this section we will focus on the properties of the mechanical spectrum $S_{cc}(\omega)$ considering both weak and strong coupling. Applying a quantum noise approach we calculate the modifications of the mechanical spectrum due to weak dissipative coupling. We derive the optical damping and the optically-induced frequency shift from the weak-coupling force spectrum and compare them to the results in case of purely dispersive coupling. To go beyond weak coupling, we then present the exact solutions of the linearized equations of motion for the general case of dispersive and dissipative coupling. Finally, we calculate the mechanical spectrum and find signatures of normal-mode splitting and an additional feature originating from the modified force spectrum.

\subsection{Quantum noise approach}\label{subsec:QNA}
Although the linearized equations of motion can be solved exactly, cf.~Eqs.~(\ref{eqSolC}) and (\ref{eqSolD}), additional insight can be gained using a weak-coupling approach. In this subsection we discuss how the mechanical spectrum $S_{cc}(\omega)$ is modified if only a small optomechanical coupling is considered. Recall that in absence of optomechanical coupling the mechanical oscillator is still coupled to the mechanical bath. Thus the oscillator is damped at a rate $\gamma$ which leads to a mean phonon number in thermal equilibrium, $n_\mathrm{th}$. Together with the resonance frequency $\omega_m$, these quantities determine the mechanical spectrum $S_{cc}(\omega)$ which is a Lorentzian of width $\gamma$ (FWHM) with a peak at $-\omega_m$ and an area of $2\pi n_\mathrm{th}$.

For small coupling we treat the influence on the mechanical oscillator as a quantum noise source, inducing transition rates between neighbouring phonon number states, $\Gamma_{n\rightarrow n+1}$ and  $\Gamma_{n\rightarrow n-1}$, given by Fermi's Golden Rule. Defining an amplification rate $\Gamma_{\uparrow}= \Gamma_{n\rightarrow n+1}/(n+1)=x_0^2 S_{FF}(-\omega_m)$ as well as a cooling rate $\Gamma_{\downarrow}=\Gamma_{n\rightarrow n-1}/n=x_0^2 S_{FF}(\omega_m)$ independent of the phonon number, both $\Gamma_\uparrow$ and $\Gamma_\downarrow$ are determined by the weak-coupling force spectrum $S_{FF}(\omega)$ \cite{cdgms2010}. These rates lead to an optically-induced damping $\gamma_\mathrm{opt} =\Gamma_\downarrow-\Gamma_\uparrow$ and a minimal phonon number $n_\mathrm{opt}=S_{FF}(-\omega_m)/\gamma_\mathrm{opt}$. In the presence of the mechanical bath and the optomechanical coupling this results in a total damping $\gamma_\mathrm{tot}=\gamma+\gamma_\mathrm{opt}$ which determines the new width of the Lorentzian describing the mechanical spectrum $S_{cc}(\omega)$. Furthermore, the additional damping leads to a steady-state mean phonon number $n_\mathrm{osc}= (\gamma n_\mathrm{th}+\gamma_\mathrm{opt}n_\mathrm{opt})/(\gamma_\mathrm{opt}+\gamma)$, thus the area of the Lorentzian is changed. Finally, optical damping affects the effective spring constant corresponding to a shift of the mechanical frequency given by $\delta\omega_m = \int d\omega S_{FF}(\omega)[1/(\omega_m-\omega)-1/(\omega_m+\omega)]/2\pi$ \cite{mccg2007}. Thus, the modifications of the mechanical spectrum due to weak optomechanical coupling can be described by the parameters $\gamma_\mathrm{opt}$, $n_\mathrm{osc}$ and $\delta\omega_m$. 

So far our considerations do not explicitly depend on the type of the coupling. However, the force spectrum $S_{FF}(\omega)$ contains this information, i.e.~its shape depends on the applied coupling. To calculate $S_{FF}(\omega)$ we use the backaction force operator \cite{egc2009}
\begin{equation}
\hat{F}x_0 = \tilde{A}\kappa(\bar{a}^*\hat{d}+\bar{a}\hat{d}^\dagger)+i\frac{\tilde{B}}{2}\sqrt{\kappa}(\bar{a}^*\hat{\xi}_\mathrm{in}-\bar{a}\hat{\xi}_\mathrm{in}^\dagger)-\Omega\frac{\tilde{B}}{2}(\hat{d}+\hat{d}^\dagger),
\end{equation}
which can be determined from the interaction part of the Hamiltonian, i.e.~the third term of Eq.~(\ref{eqHamiltonian}).
Substituting $\Omega=-i\bar{a}(i\Delta-\kappa/2)$ derived from the steady-state solution of the uncoupled Eq.~(\ref{eqClassEOMa}), the force spectrum is given by \cite{egc2009}
\begin{eqnarray}
\label{eqForceSpec}
S_{FF}(\omega)&=&\int\limits_{-\infty}^{+\infty}\frac{d\omega'}{2\pi}\langle\hat{F}(\omega)\hat{F}(\omega')\rangle_{\tilde{A}=\tilde{B}=0}\nonumber\\
&=& \kappa\left(\frac{\tilde{B}|\bar{a}|}{2x_0}\right)^2\left|1+\left(-i\Delta-\frac{\kappa}{2}+i\frac{2\tilde{A}\kappa}{\tilde{B}}\right)\chi_c(\omega)\right|^2\\
&=&\kappa\left(\frac{\tilde{B}|\bar{a}|}{2x_0}\right)^2\left|\chi_c(\omega)\right|^2\left(\omega+2\Delta-\frac{2\tilde{A}\kappa}{\tilde{B}}\right)^2.\nonumber
\end{eqnarray}
In the general case of dispersive and dissipative coupling the result is a Fano line shape which reduces to a Lorentzian in absence of dissipative coupling, i.e.~$\tilde{B}=0$. As discussed in Ref.~\cite{egc2009} the Fano line shape originates from an interference effect between the two ways of interaction with the mechanics. These two processes act as two noise sources influencing the mechanical oscillator, and lead to the two terms inside the absolute value in Eq.~(\ref{eqForceSpec}): The constant first term accounts for the direct interaction between optical bath and mechanical oscillator and represents coupling to a continuum. In contrast, the second term is filtered by the cavity response $\chi_c(\omega)=[\kappa/2-i(\omega+\Delta)]^{-1}$ and arises due to the influence of the cavity. The interference of these two contributions, the direct action of the optical bath and its cavity-mediated influence, gives rise to the Fano line shape. Purely dispersive coupling leads only to a filtered, cavity-mediated influence, i.e.~the mechanical oscillator is only influenced by a single optical noise source, and no interference can occur.

\begin{figure}
\begin{center}
\includegraphics[width=0.45\columnwidth]{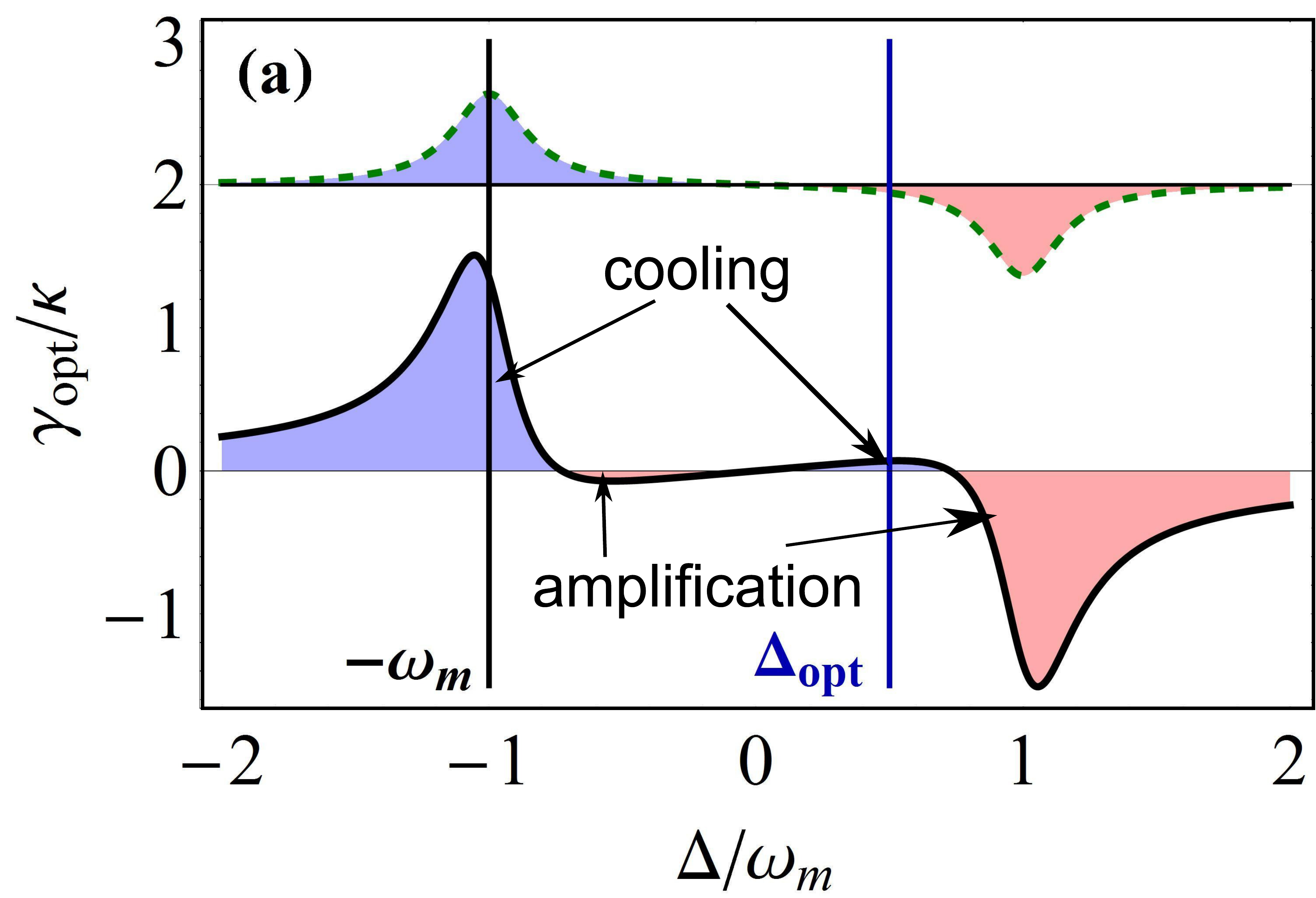}
\hspace{0.8cm}
\includegraphics[width=0.45\columnwidth]{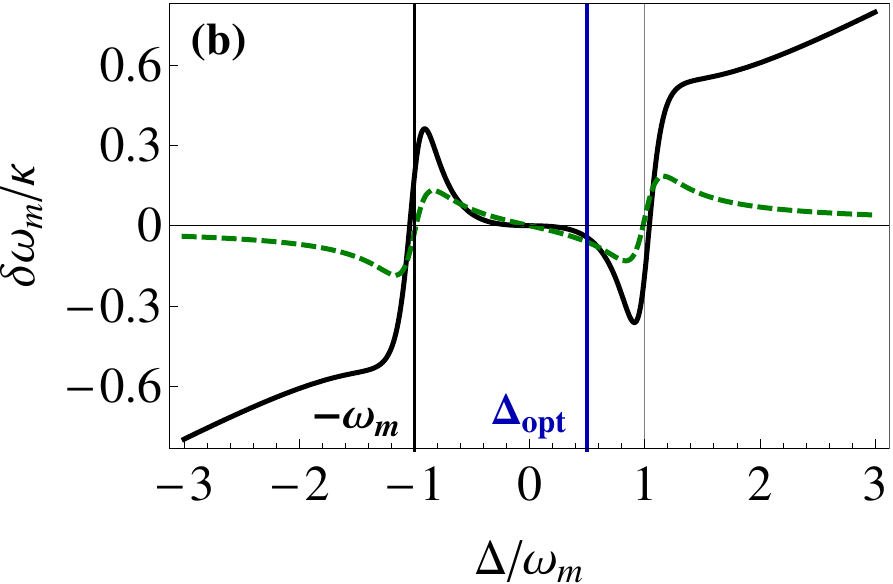}
\caption{Optical damping $\gamma_\mathrm{opt}$ (a) and optically-induced frequency shift $\delta\omega_m$ (b) as a function of detuning $\Delta$. The dashed, green lines show the result for purely dispersive coupling ($\tilde{A}\bar{a}=0.4, \tilde{B}=0$), the solid, black lines show purely dissipative coupling ($\tilde{A}=0, \tilde{B}\bar{a}=0.4$). Blue (red) areas in (a) indicate cooling (amplification). The sideband parameter is $\omega_m/\kappa=3$.}
\label{pltWeakCoupling}
\end{center}
\end{figure}

Figure \ref{pltWeakCoupling} (a) shows the optical damping for purely dispersive and purely dissipative coupling. Since $\tilde{B}=0$ means that the force spectrum is a Lorentzian, the optical damping $\gamma_\mathrm{opt}$ is given by the difference of two Lorentzians. Choosing $\Delta\approx-\omega_m$ maximizes the optical damping rate. In contrast, since the force spectrum $S_{FF}(\omega)$ is a Fano line shape for dissipative coupling (purely or in addition to dispersive coupling), the optical damping rate is modified\cite{egc2009}. The maximum is shifted farther away from the mechanical resonance frequency $\omega_m$ and for $|\Delta|\gg\kappa$ the optical damping rate decreases more slowly than a Lorentzian and is proportional  $-1/\Delta$. Furthermore, for typical parameters we find two regions where the optical damping is positive, thus providing cooling, as well as two regions with negative $\gamma_\mathrm{opt}$, leading to instability if the total damping $\gamma_\mathrm{tot}=\gamma+\gamma_\mathrm{opt}<0$.

Figure \ref{pltWeakCoupling} (b) shows the optically-induced frequency shift $\delta\omega_m$ for purely dispersive and purely dissipative coupling. Dispersive coupling and cooling at $\Delta=-\omega_m$ allows for a vanishing frequency shift $\delta\omega_m=0$. In contrast, dissipative coupling leads to a nonzero frequency shift $\delta\omega_m$ at $\Delta=-\omega_m$. It remains small for small detunings only, for large values of $\Delta$ it increases linearly. Note that this linear dependence is due to the fact that we fix the number of photons inside the cavity $|\bar{a}|^2$, which implies that the drive strength $\Omega$ has to increase with the detuning $\Delta$. Since dissipative coupling has a component proportional to $\Omega$, the effective dissipative coupling strength is increased. Fixing the laser power instead, the intra-cavity amplitude $\bar{a}$ decreases as $1/\Delta$ and $\delta\omega_m\approx(\tilde{B}|\bar{a}|)^2\Delta/2$ goes to zero in the limit of large detunings $|\Delta|\gg\omega_m$.

\begin{figure}
\begin{center}
\includegraphics[width=0.45\columnwidth]{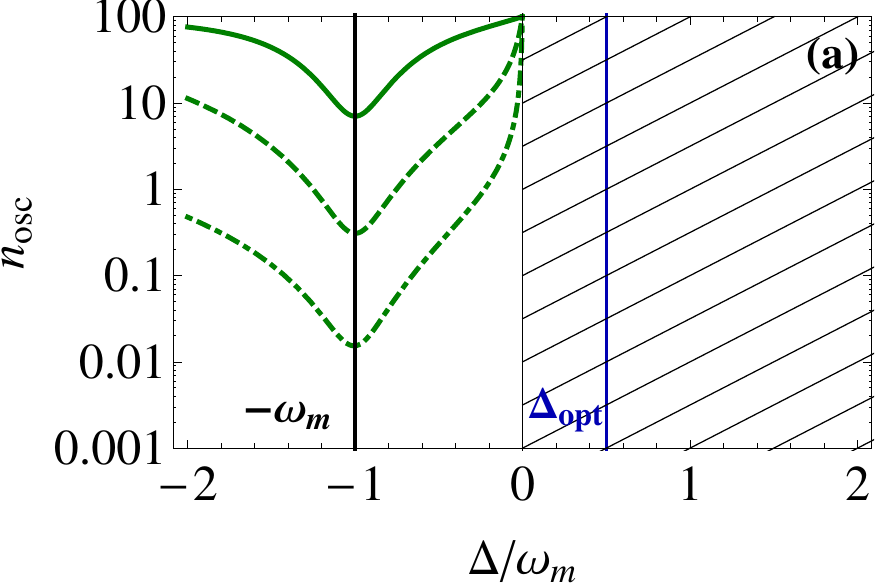}
\hspace{0.8cm}
\includegraphics[width=0.45\columnwidth]{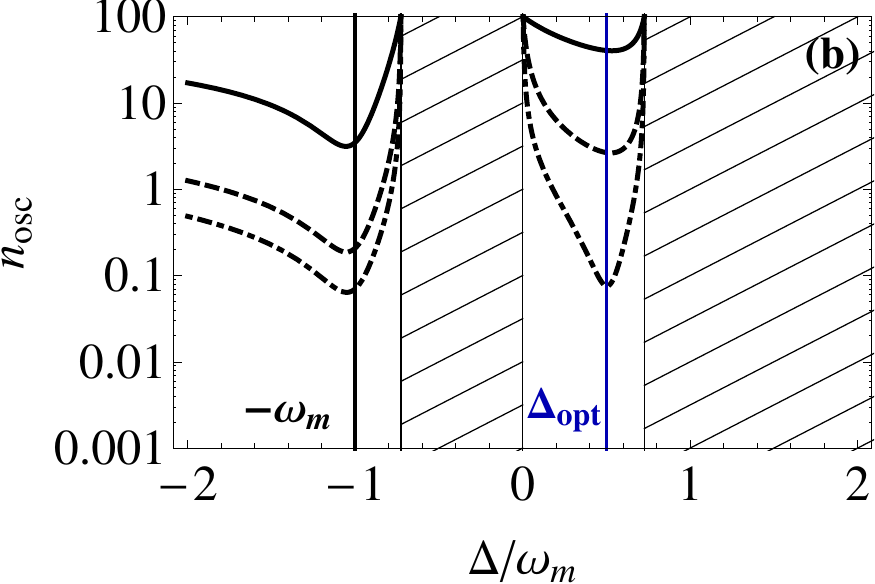}
\caption{Mean phonon number $n_\mathrm{osc}$ as a function of detuning $\Delta$  in case of (a) purely dispersive coupling and (b) purely dissipative coupling. The solid lines show the result for (a) $\tilde{A}\bar{a}=0.01$ and (b) $\tilde{B}\bar{a}=0.01$, respectively, the dashed lines show (a) $\tilde{A}\bar{a}=0.05$ and (b) $\tilde{B}\bar{a}=0.05$, and the dot-dashed lines show (a) $\tilde{A}\bar{a}=0.3$ and (b) $\tilde{B}\bar{a}=0.3$. Other parameters are $\omega_m/\kappa=3$, $\omega_m/\gamma = 10^{5}$, and $n_\mathrm{th}=100$. Hatched areas indicate unstable regions due to the criterion $\gamma_\mathrm{tot}<0$.}
\label{pltNosc}
\end{center}
\end{figure}

One possible choice to achieve cooling with dissipative coupling is $\Delta\approx-\omega_m$. Figure \ref{pltNosc} shows that, for both purely dispersive and purely dissipative coupling, $\Delta\approx-\omega_m$ leads to a strong decrease of the phonon number $n_\mathrm{osc}$. This is not surprising since the optical damping $\gamma_\mathrm{opt}$ is maximized close to this detuning in both cases. However, large optical damping alone is not sufficient to achieve the best cooling results in the sense of smallest $n_\mathrm{osc}$. Notably, dispersive coupling at this detuning leads to smaller $n_\mathrm{osc}$ despite the larger optical damping rate of dissipative coupling, except for very small coupling strengths. This is due to a larger $n_\mathrm{opt}$ which also contributes to the mean phonon number $n_\mathrm{osc}$. Thus it is of particular interest to achieve $n_\mathrm{opt}$ as small as possible. For dispersive coupling this is linked to reaching the resolved-sideband limit \cite{mccg2007}. In presence of dissipative coupling the Fano line shape of $S_{FF}(\omega)$ leads to an optimal detuning $\Delta_\mathrm{opt}=\omega_m/2+\kappa\tilde{A}/\tilde{B}$ where $n_\mathrm{opt}=0$ \cite{egc2009}. More generally, the Fano line shape of $S_{FF}(\omega)$ allows to find a detuning $\Delta_0(\omega)=-\omega/2+\kappa\tilde{A}/\tilde{B}$ for each $\omega$, such that $S_{FF}(\omega)=0$, cf.~Eq.~(\ref{eqForceSpec}). Hence, $\Delta_\mathrm{opt}\equiv\Delta_0(-\omega_m)$ implies $S_{FF}(-\omega_m)=0$ and this in turn implies $n_\mathrm{opt}=S_{FF}(-\omega_m)/\gamma_\mathrm{opt}=0$ and $\Gamma_\uparrow=0$. Therefore the optomechanical coupling induces a cooling rate $\Gamma_\downarrow$ but no amplification rate $\Gamma_\uparrow$ and ground-state cooling can be achieved if the drive strength is sufficiently large or the intrinsic damping $\gamma$ small enough. Fortunately these conditions are independent of the sideband parameter $\omega_m/\kappa$ and ground-state cooling can be performed in the unresolved-sideband regime that is easier to reach experimentally. Finally, since the optimal detuning is part of the second cooling region, $\Delta=\Delta_\mathrm{opt}$ is far from maximizing the optical damping rate, see Fig.~\ref{pltWeakCoupling} (a). Compared to the values of $\gamma_\mathrm{opt}$ achieved at $\Delta=-\omega_m$ for either dispersive or dissipative coupling, the optical damping rate at $\Delta=\Delta_\mathrm{opt}$ is rather small. Therefore, to achieve considerable cooling despite the poor cooling rate, stronger coupling or smaller intrinsic mechanical damping $\gamma$ is required, see Fig.~\ref{pltNosc} (b). 

Note that in the case of purely dissipative coupling, i.e.~$\tilde{A}=0$, the optimal detuning $\Delta_\mathrm{opt}=\omega_m/2$ corresponds to a blue detuned drive laser. On the contrary, driving a dispersively coupled system ($\tilde{B}=0$) with this detuning would lead to amplification rather than cooling.

\subsection{Strong coupling}
In this subsection we want to investigate the mechanical spectrum beyond weak coupling. Thus, we first calculate the exact solutions of Eqs.~(\ref{eqLinEOMc}) and (\ref{eqLinEOMd}) for the general case of dispersive and dissipative coupling. Then, using these results, we derive the mechanical spectrum $S_{cc}(\omega)$ and discuss its features, i.e.~compare the case of purely dissipative coupling to the case of purely dispersive coupling. 

To proceed from Eqs.~(\ref{eqLinEOMc}) and (\ref{eqLinEOMd}), we solve the uncoupled ($\tilde{A}=\tilde{B}=0$) classical equations (\ref{eqClassEOMb}) and (\ref{eqClassEOMa}) in steady-state and find $\bar{x}=0$ and $\Omega = -i\bar{a}\left(i\Delta-\frac{\kappa}{2}\right)$. Using these results and solving the coupled equations of motion (\ref{eqLinEOMc}) and (\ref{eqLinEOMd}) in Fourier space, we obtain
\begin{eqnarray}\label{eqSolC}\fl
\hat{c}(\omega)=-\frac{\sqrt{\gamma}}{\mathscr{N}(\omega)}\left\{\chi_m^{*-1}(-\omega)\hat{\eta}(\omega)-i\Sigma(\omega)\left[\hat{\eta}(\omega)+\hat{\eta}^\dagger(\omega)\right]\right\}\\
-\frac{\sqrt{\kappa}}{\mathscr{N}(\omega)}\chi_m^{*-1}(-\omega)\left[\bar{a}^*\alpha(\omega)\hat{\xi}_\mathrm{in}(\omega)-\bar{a}\alpha^*(-\omega)\hat{\xi}_\mathrm{in}^\dagger(\omega)\right], \nonumber
\end{eqnarray}
\begin{eqnarray}\label{eqSolD} \fl
\hat{d}(\omega)=-\sqrt{\kappa}\chi_c(\omega)\hat{\xi}_\mathrm{in}(\omega)+ \frac{\sqrt{\gamma}}{\mathscr{N}(\omega)}\bar{a}\left[\frac{\tilde{B}}{2}-\alpha(\omega)\right]\left[\chi_m^{*-1}(-\omega)\hat{\eta}(\omega)+\chi_m^{-1}(\omega)\hat{\eta}^\dagger(\omega)\right]\\
-\sqrt{\kappa}\frac{2i\omega_m}{\mathscr{N}(\omega)}\left[\frac{\tilde{B}}{2}-\alpha(\omega)\right]\left[|\bar{a}|^2\alpha(\omega)\hat{\xi}_\mathrm{in}(\omega)-\bar{a}^2\alpha^*(-\omega)\hat{\xi}_\mathrm{in}^\dagger(\omega)\right].\nonumber
\end{eqnarray}
Here we have used the cavity response function $\chi_c(\omega)=[\kappa/2-i(\omega+\Delta)]^{-1}$, the response of the mechanical oscillator $\chi_m(\omega)=\left[\gamma/2-i(\omega-\omega_m)\right]^{-1}$ and $\mathscr{N}(\omega)= \chi_m^{-1}(\omega)\chi_m^{*-1}(-\omega)+2\omega_m\Sigma(\omega)$. Furthermore, we have defined the optomechanical self-energy $\Sigma(\omega)$ and $\alpha(\omega)$ as 
\begin{equation}
\begin{array}{l}
\Sigma(\omega) = \Sigma_{\tilde{A}}(\omega)+\Sigma_{\tilde{B}}(\omega)+\Sigma_{\tilde{A}\tilde{B}}(\omega)\\
\alpha(\omega) = \alpha_{\tilde{A}}(\omega)+\alpha_{\tilde{B}}(\omega),
\end{array}
\end{equation}
where
\begin{equation}\label{eqSigmas}
\begin{array}{l}
\Sigma_{\tilde{A}}(\omega) =-i(\tilde{A}\kappa|\bar{a}|)^2\left[\chi_c(\omega)-\chi_c^*(-\omega)\right]\\
\Sigma_{\tilde{B}}(\omega) = i\left(\frac{\tilde{B}}{2}\right)^2\left|\bar{a}\right|^2 \left[\chi_c(\omega)\left(i\Delta+\frac{\kappa}{2}\right)^2-\chi_c^*(-\omega)\left(i\Delta-\frac{\kappa}{2}\right)^2\right]\\
\Sigma_{\tilde{A}\tilde{B}}(\omega) =\tilde{B} \tilde{A}\kappa\left|\bar{a}\right|^2\left[\chi_c(\omega)\left(i\Delta+\frac{\kappa}{2}\right)-\chi_c^*(-\omega)\left(i\Delta-\frac{\kappa}{2}\right)\right]
\end{array}
\end{equation}
and
\begin{equation}
\begin{array}{l}
\alpha_{\tilde{A}}(\omega)=i\chi_c(\omega)\tilde{A}\kappa\\
\alpha_{\tilde{B}}(\omega)=\frac{\tilde{B}}{2}-\frac{\tilde{B}}{2}\chi_c(\omega)\left(i\Delta+\frac{\kappa}{2}\right).
\end{array} 
\label{eqAlphas}
\end{equation}

Note that the Fourier transformation was applied such that $\hat{Q}^\dagger(\omega)=[\hat{Q}(-\omega)]^\dagger$ for all operators. In the purely dispersive case ($\tilde{B}=0$) the above definition of the optomechanical self-energy $\Sigma(\omega)$ reproduces the notation used in \cite{mccg2007}. Defining $\Sigma_{\tilde{B}}(\omega)$ and $\Sigma_{\tilde{A}\tilde{B}}(\omega)$ in a similar fashion, we can deduce the optical damping $\gamma_\mathrm{opt}=-2\mathrm{Im}[\Sigma(\omega_m)]$ and frequency shift $\delta\omega_m=\mathrm{Re}[\Sigma(\omega_m)]$. 
Note that this means that $\Sigma_{\tilde{B}}(\omega)$ differs from the definition in \cite{egc2009} by a factor of $-2i\omega_m \chi_m^*(-\omega)$.

Remarkably, the exact solutions of the linearized equations of motion, Eqs.~(\ref{eqSolC}) and (\ref{eqSolD}), have the same structure for both types of coupling. Apart from an additional contribution proportional to $\tilde{B}$ in Eq.~(\ref{eqSolD}), differences are hidden in the functions $\Sigma(\omega)$ and $\alpha(\omega)$. The additional dependence of  $\Sigma_{\tilde{B}}(\omega)$, $\Sigma_{\tilde{A}\tilde{B}}(\omega)$ and $\alpha_{\tilde{B}}(\omega)$ on the detuning $\Delta$ arises since, for dissipative coupling, the equations of motion (\ref{eqLinEOMc}) and (\ref{eqLinEOMd}) contain a term proportional to the drive strength $\Omega$. The constant term in $\alpha_{\tilde{B}}(\omega)$ is due to the direct interaction between the optical bath and the mechanical mode.

Finally, the fluctuations of the optical output are obtained by using $\hat{a}_\mathrm{out}=(\bar{a}_\mathrm{out}+\hat{\xi}_\mathrm{out})e^{-i\omega_d t}$ and linearizing the input-output relation (\ref{eqIO}), $\hat{\xi}_\mathrm{in}-\hat{\xi}_\mathrm{out}=-\sqrt{\kappa}\hat{d}-\sqrt{\kappa}\tilde{B}\bar{a}\hat{x}/2x_0$.
Then, with Eqs.~(\ref{eqSolC}) and (\ref{eqSolD}), we find 
\begin{eqnarray}\label{eqSolOut}\fl
\hat{\xi}_\mathrm{out}(\omega) =-\frac{\sqrt{\kappa\gamma}}{\mathscr{N}(\omega)}\bar{a}\alpha(\omega)\left[\chi_m^{*-1}(-\omega)\hat{\eta}(\omega)+\chi_m^{-1}(\omega)\hat{\eta}^\dagger(\omega)\right]\\ +\left[1-\kappa\chi_c(\omega)\right]\hat{\xi}_\mathrm{in}(\omega)+\frac{2i\kappa\omega_m\alpha(\omega)}{\mathscr{N}(\omega)}\left[|\bar{a}|^2\alpha(\omega)\hat{\xi}_\mathrm{in}(\omega)-\bar{a}^2\alpha^*(-\omega)\hat{\xi}_\mathrm{in}^\dagger(\omega)\right].\nonumber
\end{eqnarray}

Using Eq.~(\ref{eqSolC}) we can calculate the mechanical spectrum 
\begin{eqnarray}\label{eqScc}
S_{cc}(\omega) = \int\limits_{-\infty}^{+\infty}\frac{d\omega'}{2\pi}\langle\hat{c}^\dagger(\omega)\hat{c}(\omega')\rangle 
= \frac{\gamma \sigma_\mathrm{th}(\omega)+\kappa\sigma_\mathrm{opt}(\omega)}{\left|\mathscr{N}(\omega)\right|^2},
\end{eqnarray}
where $\sigma_\mathrm{th}(\omega) = \left|\Sigma(\omega)\right|^2(n_\mathrm{th}+1)+\left|\chi_m^{-1}(\omega)+i\Sigma(\omega)\right|^2 n_\mathrm{th}$ and $\sigma_\mathrm{opt}(\omega) =\left|\chi_m^{-1}(\omega)\right|^2|\bar{a}|^2|\alpha(\omega)|^2=\left|\chi_m^{-1}(\omega)\right|^2S_{FF}(\omega)x_0^2/\kappa$.
This result is valid for purely dispersive, purely dissipative and both types of coupling but has the same form as found in the case of dispersive coupling only \cite{mccg2007}. 
For $\tilde{B}=0$ the result coincides with \cite{mccg2007}; setting $\tilde{A}=0$ the result coincides with \cite{egc2009}.

\begin{figure}[tbp]
\centering
\includegraphics[width=1\columnwidth]{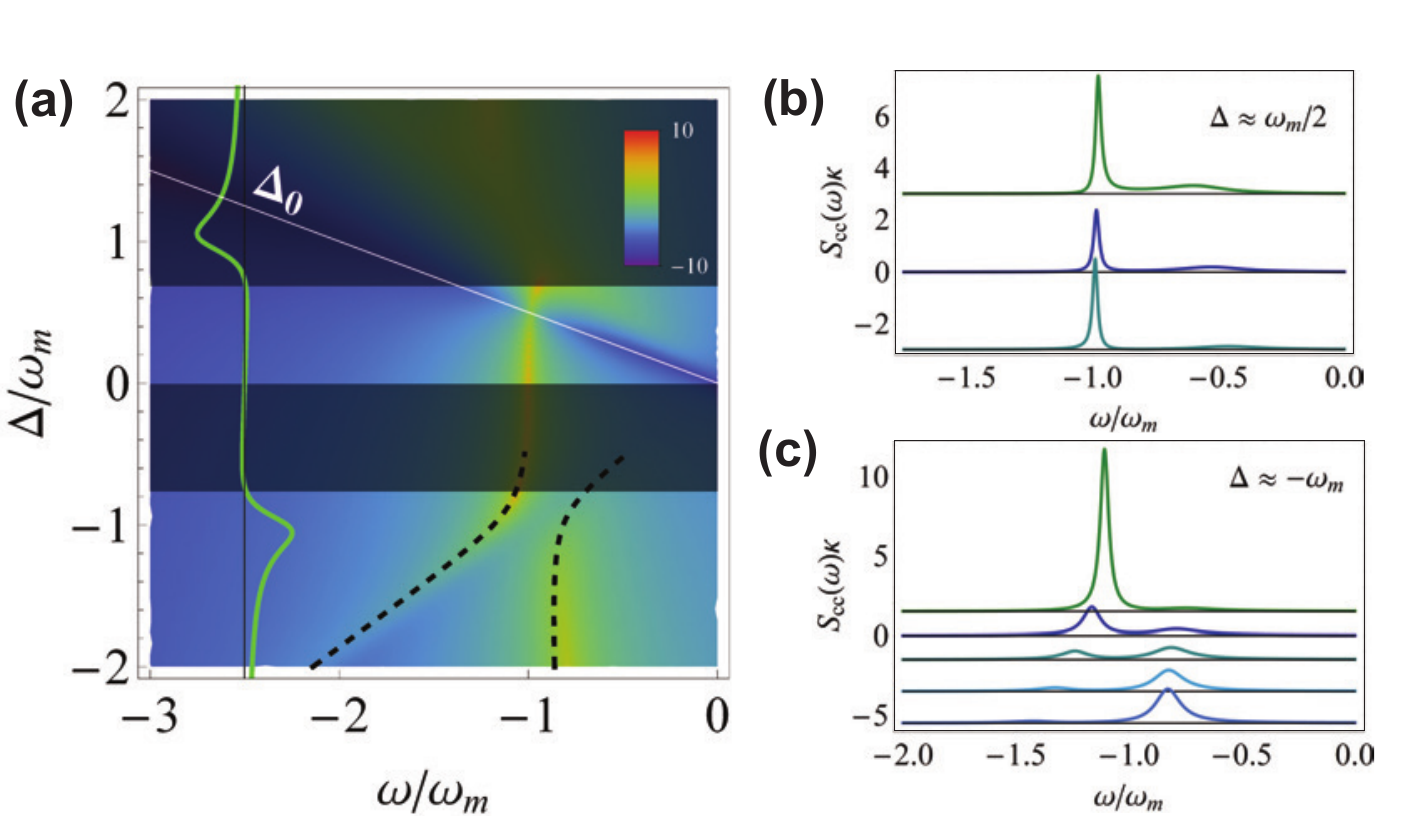}
\caption{(a) Logarithm of the mechanical spectrum $S_{cc}(\omega)\kappa$ as a function of detuning $\Delta$. Parameters are $\omega_m/\kappa=3$, $\omega_m/\gamma = 10^{5}$, $n_\mathrm{th}=100$, $\tilde{A}=0$, and $\tilde{B}\bar{a}=0.4$. The dark regions indicate regions of instability obtained from a numerical calculation. The green curve gives half of the total damping rate, $\gamma_\mathrm{tot}/2$, obtained from the quantum noise approach with the origin shifted to $(-2.5,0)$. The dashed lines show the real part of the eigenvalues of the Hamiltonian (\ref{eqHsimp}). (b) $S_{cc}(\omega)\kappa$ for detunings $\Delta/\omega_m=0.55,0.5,0.45$ (from top to bottom). (c) $S_{cc}(\omega)\kappa$ for detunings $\Delta/\omega_m=-0.9,-1,-1.1,-1.2,-1.3$ (from top to bottom).
}
\label{pltSccDensity}
\end{figure}

Figure \ref{pltSccDensity} (a) shows the mechanical spectrum for strong dissipative coupling. Dark areas indicate regions where the solutions of the linearized equations of motion are unstable. This was numerically tested for the parameters used in Fig.~\ref{pltSccDensity} and coincides with the regions where the total damping rate $\gamma_\mathrm{tot}$ from the weak-coupling approach is negative. Whereas dispersive coupling leads to one unstable region for blue detuning, dissipative coupling can lead to a second unstable region for red detuning in addition to an unstable region for blue detuning. A third unstable region exists for even stronger drive or large red detuning. This is not predicted by the behaviour of the optical damping rate, i.e.~it appears although $\gamma_\mathrm{tot}>0$.

Focusing on the stable regions, we find two prominent features. First, at $\Delta=\Delta_\mathrm{opt}=\omega_m/2$ a strong decrease of the phonon number $\langle \hat{n}\rangle=\int d\omega S_{cc}(\omega)/2\pi $ can be observed. As mentioned in Subsection \ref{subsec:QNA}, this detuning is associated with cooling \cite{egc2009} and a special case of the strong modifications of the mechanical spectrum at $\Delta_{0}(\omega)$. If $\Delta=\Delta_0(\omega)$, the force spectrum $S_{FF}(\omega)$ vanishes, which means that at this frequency $\omega$ only the first term of Eq.~(\ref{eqScc}), $\sigma_\mathrm{th}(\omega)$, contributes to $S_{cc}(\omega)$. Furthermore, Fig.~\ref{pltSccDensity} (b) shows that, apart from the main peak close to the mechanical resonance $\omega=-\omega_m$, there is a broad contribution at a second frequency arising from $\sigma_\mathrm{opt}(\omega)/|\mathscr{N}(\omega)|^2$. It is given as a trade-off between the maximum of the Fano line shape of the force spectrum $S_{FF}(\omega)$ at $\omega=\frac{-4\Delta^2+\kappa ^2}{4\Delta}$ and the peak of $|\mathscr{N}(\omega)|^{-2}$ at $\omega=-\omega_m$. It is this contribution, away from the mechanical resonance frequency, that finally limits the cooling due to its increasing relevance with increasing coupling strength. 

The second feature is found at $\Delta=-\omega_m$. Similar to the case of dispersive coupling, we find normal-mode splitting even though slight quantitative differences appear. In the following we use a simplified Hamiltonian to find an approximation that describes the splitting. 
Recall that dissipative coupling leads to two terms in the equations of motion (\ref{eqLinEOMc}) and (\ref{eqLinEOMd}). We neglect the term proportional to the damping rate, i.e.~the direct influence of the optical bath on the mechanical oscillator, and only take the effect proportional to the drive into account. Furthermore, we use the rotating wave approximation and neglect the fast rotating terms $\hat{d}^\dagger\hat{c}^\dagger$ and $\hat{d}\hat{c}$. Then, in the rotating frame, the simplified, non-hermitian Hamiltonian is given by
\begin{equation}\label{eqHsimp}\fl
\hat{H}= -\left(\Delta+i\frac{\kappa}{2}\right)\hat{d}^\dagger\hat{d}+\left(\omega_m-i\frac{\gamma}{2}\right)\hat{c}^\dagger\hat{c}+\left[\left(\frac{\tilde{B}\Omega}{2}-\tilde{A}\bar{a}\kappa\right)\hat{c}\hat{d}^\dagger+\mathrm{H.c.}\right].
\end{equation}
Note that using this approximation, the difference between purely dispersive and purely dissipative coupling only depends on whether the drive strength $\Omega$ or the intra-cavity amplitude $\bar{a}$ is fixed. Fixing $\Omega$ for purely dissipative and $\bar{a}$ for purely dispersive coupling leads to similar results. Instead fixing one parameter for both types of coupling, as done here with a variable drive strength $\Omega$ and a fixed $\bar{a}$, leads to modifications of the splitting due to an additional dependence on the detuning $\Delta$. Since it is not possible to fix both $\Omega$ and $\bar{a}$ at the same time, mixed coupling will always lead to $\Delta$-dependent modifications arising from either the dissipative or the dispersive term.

In the general case of dispersive and dissipative coupling, the eigenvalues of the simplified Hamiltonian (\ref{eqHsimp}) can be calculated as
\begin{eqnarray}\fl\label{eqEV}
E_\pm=-i\frac{\gamma+\kappa}{4}+\frac{\omega_m-\Delta}{2}\\
\pm\sqrt{-\left[\gamma-\kappa+2i\left(\Delta+\omega_m\right)\right]^2+|\bar{a}|^2\left[16(\tilde{A}^2\kappa^2-\tilde{A}\tilde{B}\Delta\kappa)+\tilde{B}^2\left(4\Delta^2+\kappa^2\right)\right]}.\nonumber
\end{eqnarray}
The energies corresponding to the two modes are the real parts of these eigenvalues $E_\pm$, whereas the imaginary parts contain information about the associated linewidths. We show the real parts of the eigenvalues (calculated for $\tilde{A}=0$) in Fig.~\ref{pltSccDensity} (a) and, despite the simplifications, the energies fit the peak position of the spectrum very well.
Differences to purely dispersive coupling arise since the dispersive coupling matrix element is constant for fixed values of the cavity amplitude $\bar{a}$. In contrast, the dissipative coupling matrix element depends on the drive strength $\Omega$, which is a function of detuning $\Delta$ if $\bar{a}$ is fixed. This affects the curvature of the modes and leads to $\Delta$-dependent width of the splitting. Moreover, the eigenvalues of the simplified Hamiltonian indicate that, in case of purely dissipative coupling, the splitting is no longer minimal at $\Delta=-\omega_m$. Neglecting the damping terms in the Hamiltonian (\ref{eqHsimp}) the minimal splitting occurs at $\Delta=-\omega_m/(1+\tilde{B}^2|\bar{a}|^2)$. Figure \ref{pltSccDensity} (c) shows in detail how the single peak at the mechanical frequency is split due to the optomechanical coupling.

\begin{figure}[tbp]
\centering
\includegraphics[width=0.49\columnwidth]{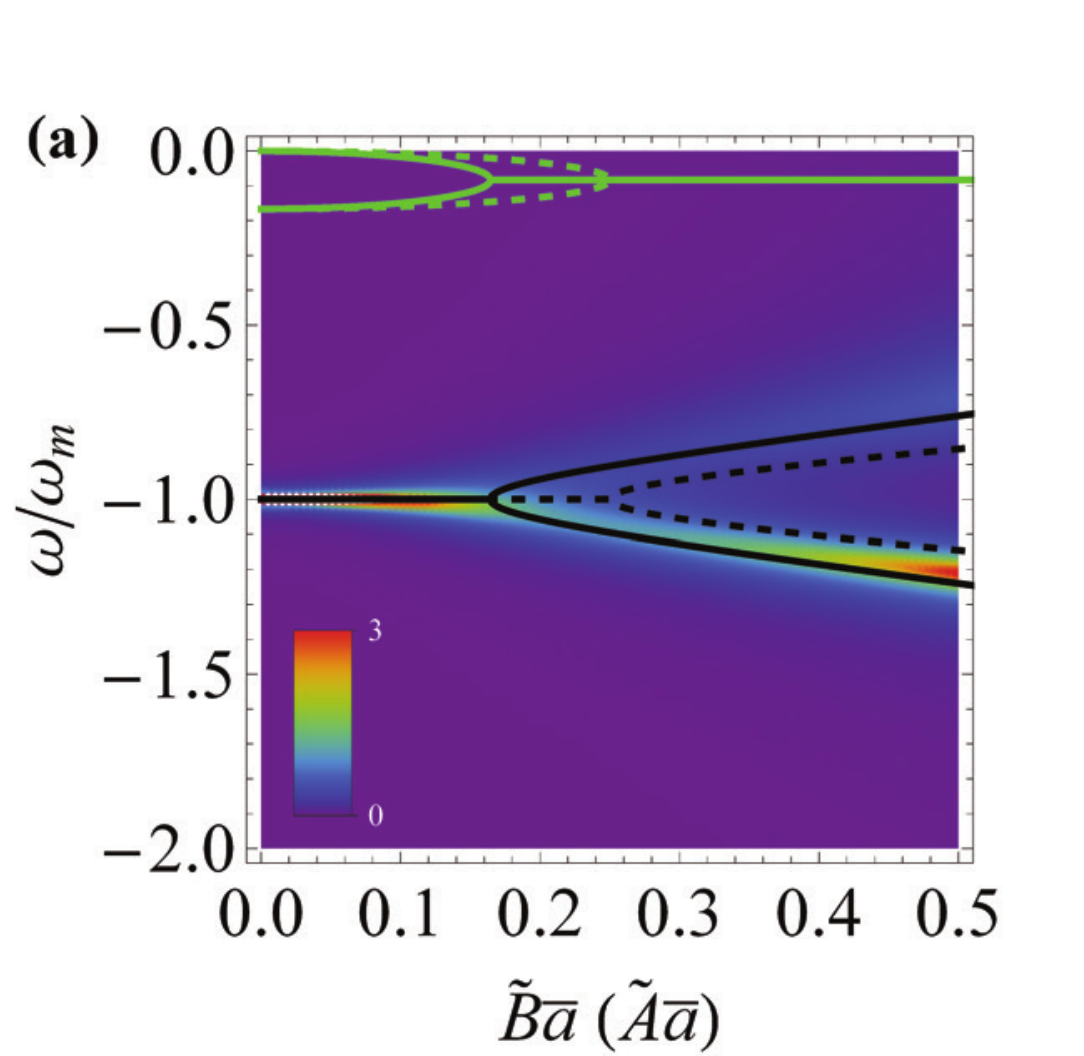}
\includegraphics[width=0.49\columnwidth]{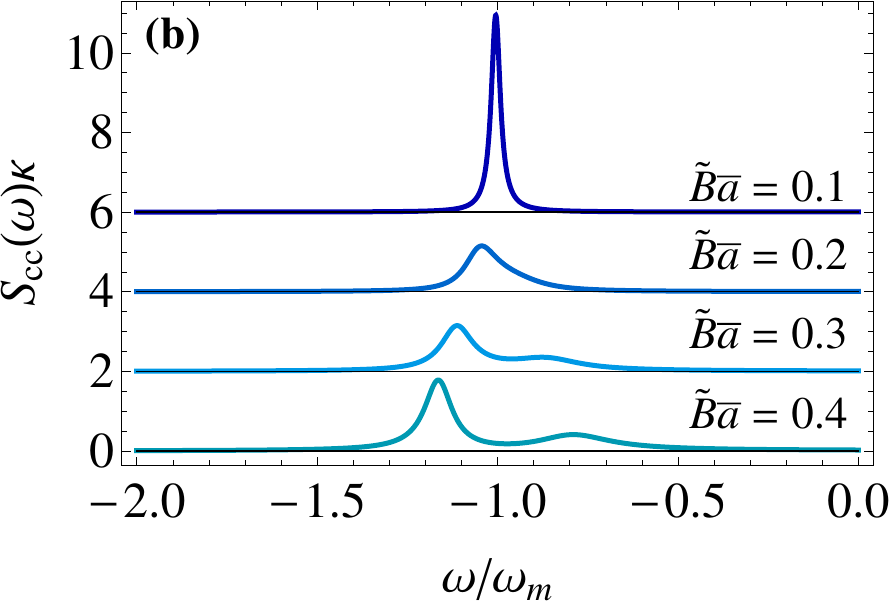}
\caption{(a) Real part (black curves) and imaginary part (green curves) of the eigenvalues $E_\pm$ calculated from the Hamiltonian (\ref{eqHsimp}) as a function of the coupling strength. Solid (dashed) lines indicate purely dissipative (dispersive) coupling. The plot is underlaid with the mechanical spectrum $S_{cc}(\omega)\kappa$ as a function of coupling strength $\tilde{B}\bar{a}$ for $\Delta=-\omega_m$ and $\tilde{A}=0$. Other parameters are $\omega_m/\kappa=3$, $\omega_m/\gamma =10^{5}$, and $n_\mathrm{th}=100$. (b) Mechanical spectrum $S_{cc}(\omega) \kappa$ at $\Delta=-\omega_m$ for different coupling strengths $\tilde{B}\bar{a}$ as well as $\tilde{A}=0$.}
\label{pltEV}
\end{figure}

We further investigate the eigenvalues $E_\pm$ from Eq.~(\ref{eqEV}) to clarify at which coupling strength normal-mode splitting appears in the mechanical spectrum $S_{cc}(\omega)$. Figure \ref{pltEV} (a) shows that $\mathrm{Re}[E_\pm]$ coincide well with the peak positions of the mechanical spectrum as a function of coupling strength $\tilde{B}\bar{a}$. If $\Delta=-\omega_m$, small coupling corresponds to degenerate energies, i.e.~$\mathrm{Re}[E_+]=\mathrm{Re}[E_-]$. In this case, the argument of the square root in Eq.~(\ref{eqEV}) is real and negative. Thus, for small coupling, the root contributes only to the imaginary part of the eigenvalues and affects the linewidths given by $\kappa$ and $\gamma$ respectively. With increasing coupling strength the linewidths approach their mean value $(\kappa+\gamma)/2$, which is reached where the root becomes zero. Then the modes $\mathrm{Re}[E_\pm]$ start to split whereas the linewidths remain unchanged. In case of purely dissipative coupling at $\Delta=-\omega_m$, the critical coupling strength where mode-splitting starts is given by $\tilde{B}|\bar{a}|=(\kappa-\gamma)/\sqrt{4\omega_m^2+\kappa^2}$. 

Figure \ref{pltEV} (a) shows normal-mode splitting in case of purely dispersive coupling. For the set of parameters used, it starts at a larger coupling strength than the splitting obtained for purely dissipative coupling. Note, however, that this depends on the sideband parameter $\omega_m/\kappa$, since the critical dispersive coupling strength at $\Delta=-\omega_m$ is given by $\tilde{A}|\bar{a}|=(\kappa-\gamma)/(4\kappa)$. Thus, for $\omega_m^2/\kappa^2<15/4$, dispersive coupling would lead to normal-mode splitting at a smaller coupling strength than dissipative coupling.

Finally note that if $\Delta\neq-\omega_m$, the root in Eq.~(\ref{eqEV}) is complex valued and the modes start with a finite energy separation from the uncoupled case.

\section{Optical output spectrum} \label{sec:optical}

The optical spectra, especially the optical output spectrum, are experimentally easier accessible than the mechanical spectrum. Thus we use the full solutions (\ref{eqSolD}) and (\ref{eqSolOut}) to calculate the cavity and the optical output spectrum. Purely dispersive coupling allows interaction between the mechanical element and the optical output only via the cavity, such that $S_{dd}^\mathrm{out}(\omega) = \kappa S_{dd}(\omega)$. Note that this is no longer the case for dissipative coupling since there is direct influence of the mechanical oscillator on the output which is not mediated by the cavity. We find
\begin{equation} \fl\label{eqCavitySpec}
S_{dd}(\omega) = \frac{|\bar{a}|^2|\alpha(-\omega)-\tilde{B}/2)|^2}{|\mathscr{N}(\omega)|^2}\left[4\kappa|\bar{a}|^2\omega_m^2|\alpha(\omega)|^2+\gamma|\chi_m^{-1}(-\omega)|^2(n_\mathrm{th}+1)+\gamma|\chi_m^{-1}(\omega)|^2n_\mathrm{th}\right]
\end{equation}
and
\begin{equation}
S_{dd}^\mathrm{out}(\omega) =\kappa\frac{|\alpha(-\omega)|^2}{|\alpha(-\omega)-\tilde{B}/2|^2}S_{dd}(\omega).
\end{equation}
Apart from the factor $\kappa$, these two spectra differ by the subtraction of the constant term from $\alpha(-\omega)$. Recalling the definition of $\alpha(\omega)$, we find that this means, that dissipative coupling contributes to the cavity spectrum only at frequencies filtered by the cavity response $|\chi_c(-\omega)|^2$. This leads to the enhancement of the lower sideband for $\Delta<0$ and of the upper sideband if $\Delta>0$, similar to the case of dispersive coupling. Due to the direct influence of the mechanical oscillator on the optical output, dissipative coupling leads to a contribution to the output spectrum $S_{dd}^\mathrm{out}(\omega)$ that is not filtered by the cavity response. This is hidden in the definition of $\alpha(-\omega)$ in Eq.~(\ref{eqAlphas}). 

\begin{figure}[tbp]
\centering
\includegraphics[width=0.495\textwidth]{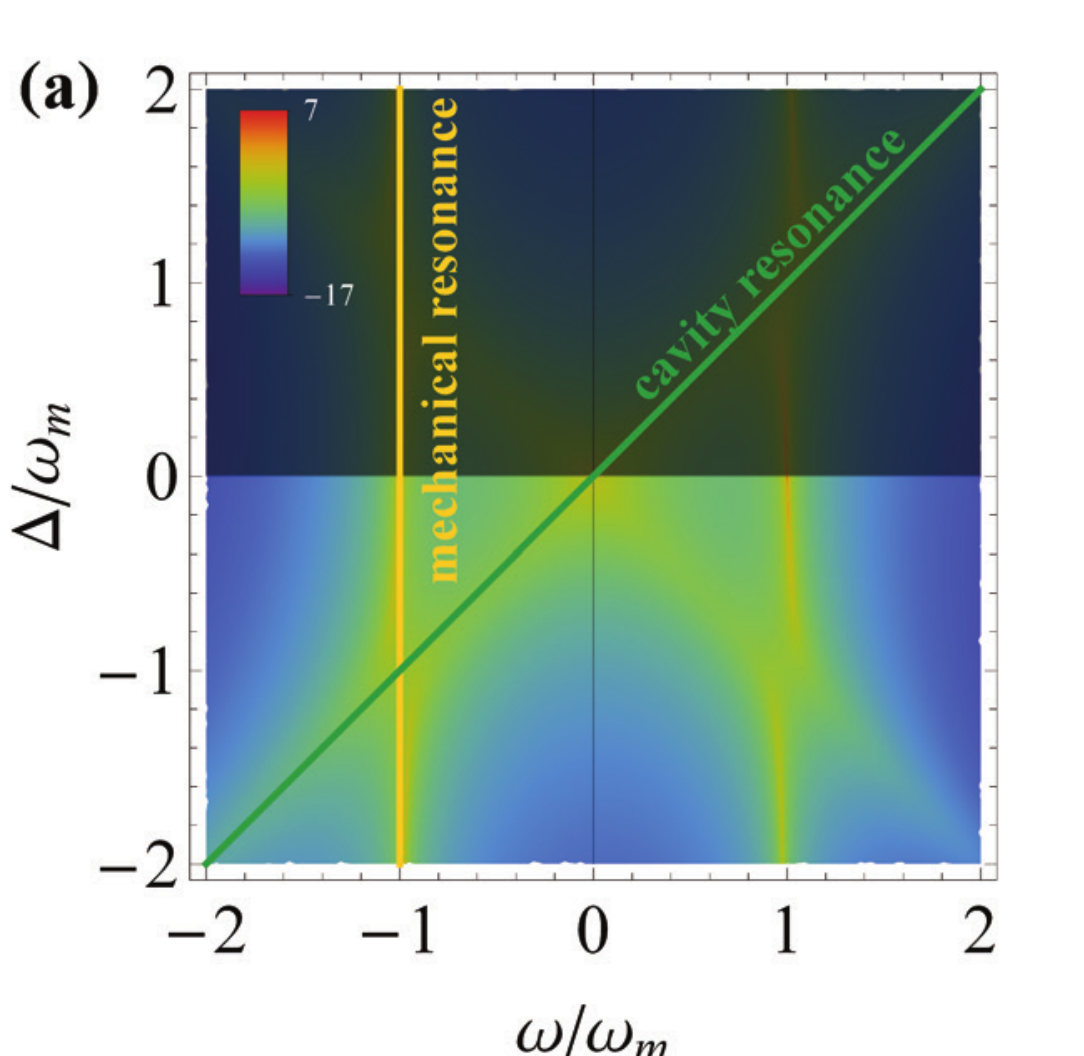}
\includegraphics[width=0.495\textwidth]{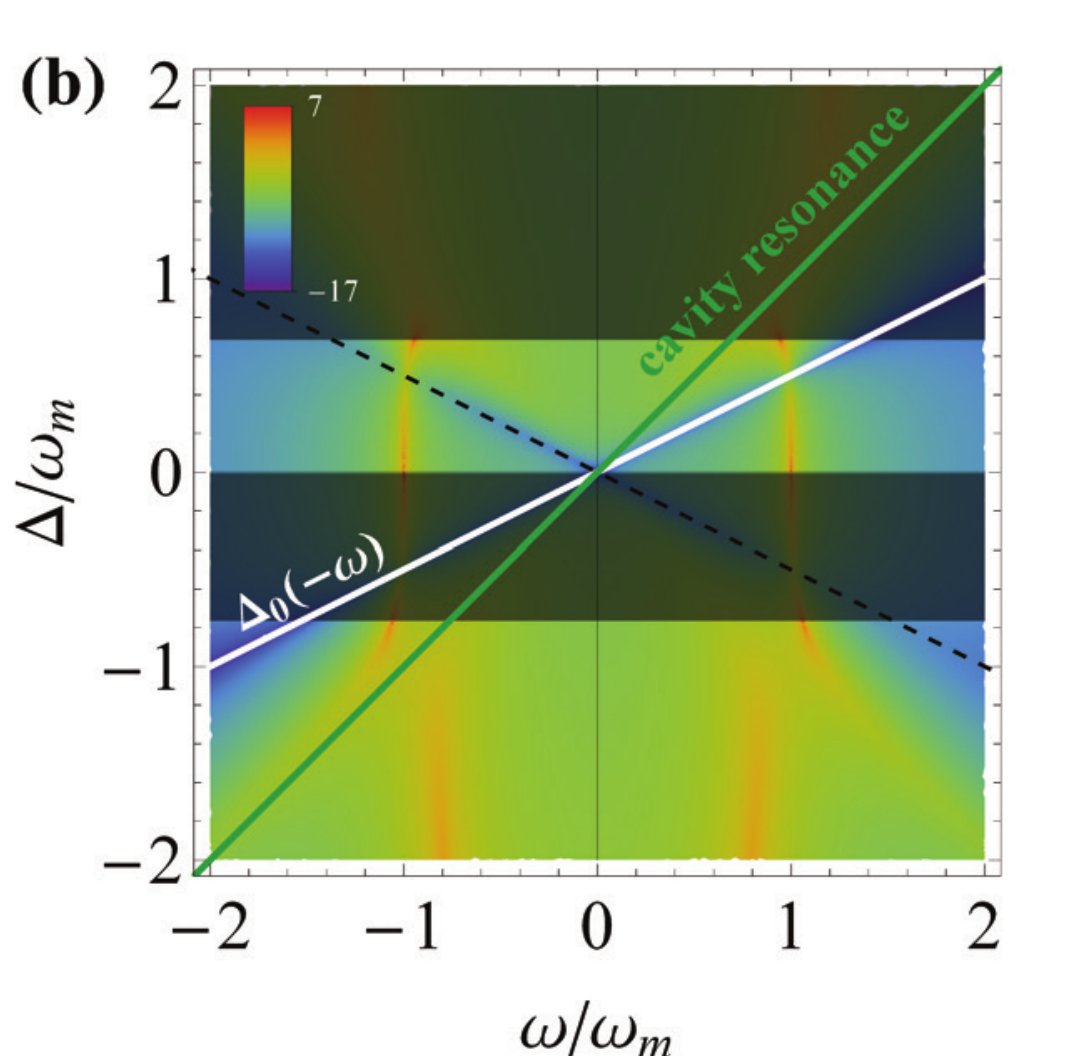}
\caption{Logarithm of the optical output spectrum $S_{dd}^\mathrm{out}(\omega)$ for (a) purely dispersive coupling ($\tilde{A}\bar{a}=0.4$, $\tilde{B}=0$) and (b) purely dissipative coupling ($\tilde{A}=0$, $\tilde{B}\bar{a}=0.4$). Other parameters are the same as in Fig.~\ref{pltSccDensity}. Dark regions indicate regions of instability. The white line indicates an exact zero of $S_{dd}^\mathrm{out}(\omega)$. The dashed line indicates where the mechanical oscillator experiences dissipative cooling associated with the white line in Fig.~\ref{pltSccDensity} (a). 
}
\label{pltDensitiesOut}
\end{figure}

The optical output spectrum is connected to the displacement spectrum $S_{xx}(\omega)$ via $S_{dd}^\mathrm{out}(\omega)=S_{FF}(-\omega)S_{xx}(\omega)$\cite{egc2009}. 
Thus, it is possible to observe the features of the mechanical spectrum $S_{cc}(\omega)\kappa$ in the optical output spectrum. As shown in Fig.~\ref{pltDensitiesOut} (b) we recover normal-mode splitting at $\Delta=-\omega_m$ and find modifications of the optical output spectrum $S_{dd}^\mathrm{out}(\omega)$ for $\Delta=\Delta_0(\pm\omega)$. First, we can see the influence of $\Delta_{0}$ on the mechanical spectrum $S_{cc}(\omega)$ at $\Delta=\Delta_0(+\omega)$. Moreover, there is also the direct influence through the weak-coupling force spectrum $S_{FF}(-\omega)$, i.e.~the optical output spectrum becomes exactly zero if $\Delta=\Delta_0(-\omega)$. 
Figure \ref{pltDensitiesOut} (a) shows $S_{dd}^\mathrm{out}(\omega)$ in case of purely dispersive coupling for comparison. Normal-mode splitting can be observed as well, but for purely dispersive coupling the detuning $\Delta_0$ has no special role. Note also the different instability regions of the optical output spectrum, depending on the type of coupling.

For mixed coupling (i.e.~$\tilde{A}\neq 0$ and $\tilde{B}\neq 0$) the new features of dissipative coupling are modified but do not disappear. In particular, there is a detuning $\Delta_{0}(\omega)$ such that $S_{FF}(\omega)=0$. However, its offset $\kappa\tilde{A}/\tilde{B}$ depends on the ratio of the couplings and leads to a shift of $\Delta_0$ compared to the purely dissipative case. Furthermore, mixed coupling modifies the regions where $\gamma_\mathrm{tot}<0$ in the weak-coupling approach and the corresponding changes of the unstable regions are captured by the numerical calculation as well.

\section{Optomechanically-induced transparency (OMIT)} \label{sec:OMIT}

In this section we will investigate the response of the optomechanically-coupled system to a weak probe field, and we show that purely dissipative coupling, i.e.~$\tilde{A}=0$, leads to optomechanically-induced transparency. This is also a convenient way to observe normal-mode splitting (NMS). We compare our findings to the purely dispersive case and give an appropriate approximation that holds in the general case of dispersive and dissipative coupling, i.e.~$\tilde{A}\neq 0$ and $\tilde{B}\neq 0$, in the resolved-sideband regime.

The probe field of frequency $\omega_p$ is assumed to be weak compared to the drive field, i.e.~its optomechanical coupling can be neglected.
Thus it is sufficient to account for the probe laser by changing the optical input mode $\hat{\xi}_\mathrm{in}(\omega)$ in an appropriate way and neglecting additional coupling terms. In Section \ref{sec:mechanical} and \ref{sec:optical} the operator $\hat{\xi}_\mathrm{in}$ denoted vacuum fluctuations only, now it contains the probe field such that $\hat{\xi}_\mathrm{in}(t)=\hat{\xi}_\mathrm{vac}(t)+\bar{d}_\mathrm{probe}e^{-i\delta t}$ with $\langle\hat{\xi}_\mathrm{in}(t)\rangle=\bar{d}_\mathrm{probe}e^{-i\delta t}$. Here $\hat{\xi}_\mathrm{vac}$ describes the vacuum fluctuations of the optical bath and $\delta=\omega_p-\omega_d$ denotes the detuning between probe and drive laser.

We investigate the response to the probe field by evaluating the expectation value of the optical output mode (\ref{eqSolOut}). Since $\langle\hat{\eta}\rangle=\langle\hat{\eta}^\dagger\rangle=0$, the result is of the form
\begin{equation}
\label{eqXiOut}
\langle\hat{\xi}_\mathrm{out}(t)\rangle=\int_{-\infty}^{+\infty} \frac{d\omega}{2\pi} \langle\hat{\xi}_\mathrm{out}(\omega)\rangle e^{-i\omega t}=A^-e^{-i\delta t}+A^+e^{i\delta t}.
\end{equation}
Recall that all calculations are done in a frame rotating with $-\omega_d$, thus the optical output contains terms rotating at three frequencies: $-\omega_d$ (drive frequency), $-\delta-\omega_d=-\omega_p$ (anti-Stokes field) and $\delta-\omega_d=\omega_p-2\omega_d$ (Stokes field). 
The contribution at the drive frequency is not contained in Eq.~(\ref{eqXiOut}) since we treated the coherent part of the drive separately with Eqs.~(\ref{eqClassEOMb}) and (\ref{eqClassEOMa}), i.e.~$\hat{\xi}_\mathrm{out}$ only describes the fluctuations around the strong drive field. $A^-$ and $A^+$ are the complex amplitudes of the anti-Stokes and Stokes field and are given by

\begin{equation}\label{eqA-}
A^-=\left[1-\kappa\chi_c(\delta)+2i\kappa\omega_m\frac{|\bar{a}|^2\alpha(\delta)^2}{\mathscr{N}(\delta)}\right]\bar{d}_{\mathrm{probe}},
\end{equation}

\begin{equation}\label{eqA+}
A^+=-2i\kappa\omega_m\frac{\bar{a}^2\alpha^*(\delta)\alpha(-\delta)}{\mathscr{N}(-\delta)}\bar{d}_{\mathrm{probe}}^*.
\end{equation}
The anti-Stokes field rotates with the probe frequency $-\omega_p$, thus $A^-$ is the amplitude of the original probe field modified due to interference with anti-Stokes scattered light ($\delta>0$) from the drive field. Furthermore, $A^+$ is the amplitude of the output field component rotating at a frequency $\omega_p-2\omega_d$ that is created by the optomechanical coupling, i.e.~Stokes scattering ($\delta>0$) of drive photons. 

\begin{figure}[tbp]
\includegraphics[width=0.45\columnwidth]{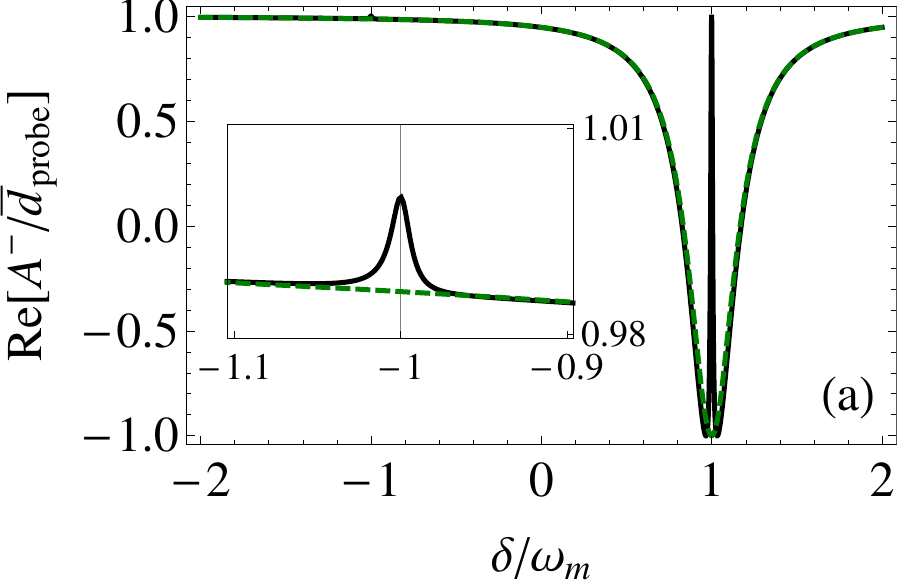}\hspace{0.1\columnwidth}
\includegraphics[width=0.45\columnwidth]{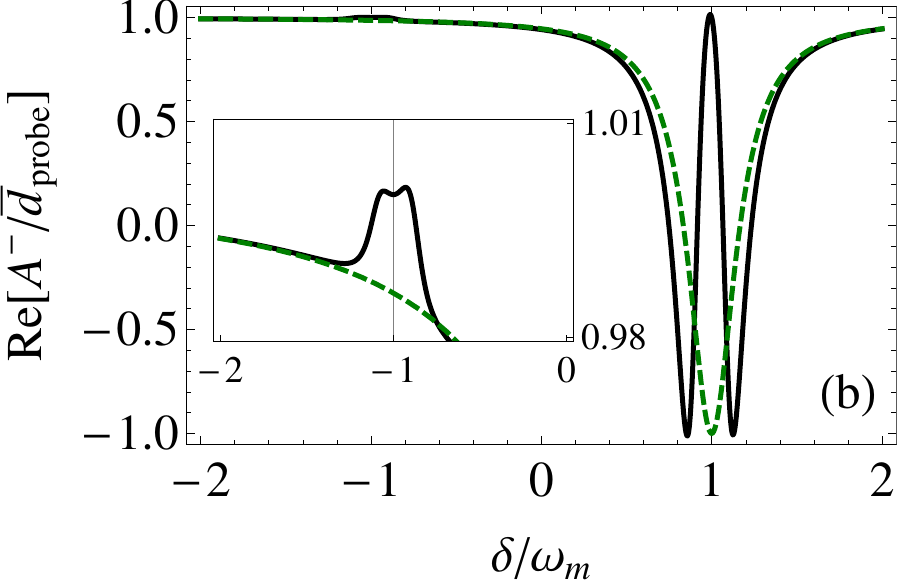}\\
\includegraphics[width=0.45\columnwidth]{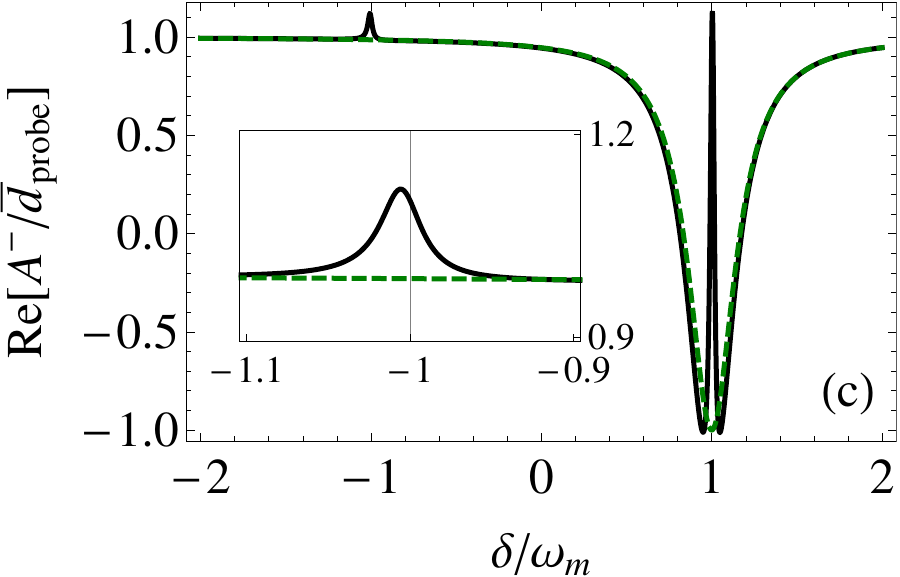}\hspace{0.1\columnwidth}
\includegraphics[width=0.45\columnwidth]{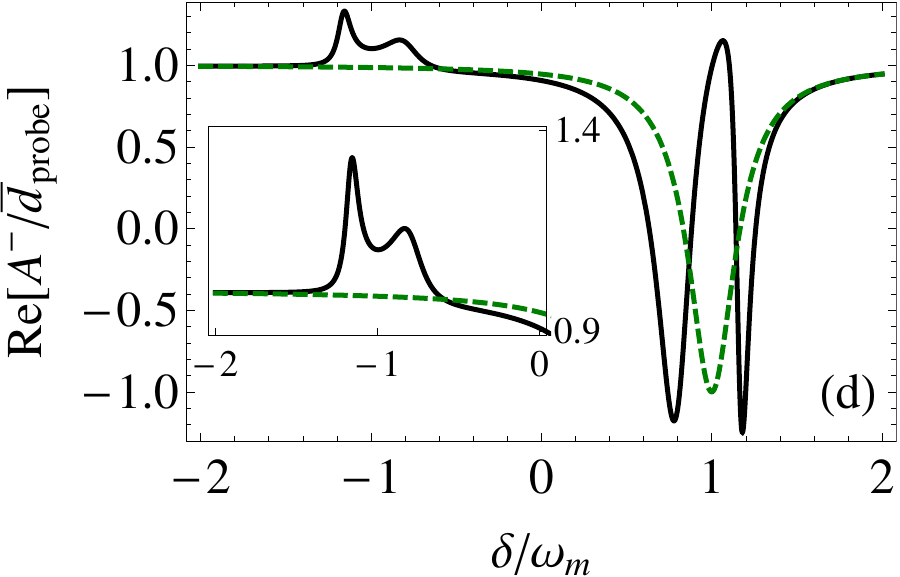}
\caption{The real part $\mathrm{Re}[A^-/\bar{d}_\mathrm{probe}]$ of the response at the probe frequency $-\omega_p$ as a function of the detuning between probe and drive field $\delta$ for $\Delta=-\omega_m$. (a) and (b) show the case of purely dispersive coupling with $\tilde{A}\bar{a}=0.1$  and $\tilde{A}\bar{a}=0.4$. (c) and (d) show purely dissipative coupling with $\tilde{B}\bar{a}=0.1$ and $\tilde{B}\bar{a}=0.4$. Other parameters are the same as in Fig.~\ref{pltSccDensity}. The green dashed line shows the result in absence of coupling ($\tilde{A}=\tilde{B}=0$). The insets show a magnification around $\delta=-\omega_m$.}
\label{pltOMIT}
\end{figure}

Focusing on the anti-Stokes contribution at $\Delta=-\omega_m$ where NMS appears in the mechanical spectrum $S_{cc}(\omega)$, Eq.~(\ref{eqA-}) consists of three contributions to the amplitude $A^-$: The constant first term accounts for the initial probe field. The second term represents the influence of the uncoupled cavity. Finally, the third term is nonzero only for nonzero coupling and contains the influence of both dispersive and dissipative coupling. 

Using homodyne detection, different quadratures of the anti-Stokes field can be investigated experimentally. Figure \ref{pltOMIT} shows the real part of the anti-Stokes amplitude, $\mathrm{Re}[A^-]$. In absence of optomechanical coupling the cavity leads to a Lorentzian-shaped dip of width $\kappa$ associated with the second term in Eq.~(\ref{eqA-}). For nonzero coupling the third term in Eq.~(\ref{eqA-}) modifies the amplitude due to scattering processes from the drive to this frequency. These processes are suppressed away from the mechanical resonance, thus striking modifications occur only for $\delta\approx\pm\omega_m$. There, an upper or lower sideband $-\omega_d\pm\omega_m$ is created and its frequency coincides with the probe frequency $-\omega_p$, which gives rise to interference effects \cite{wrdgask2010,t2011NMS}.

At $\delta=-\omega_m$, scattering from the drive laser is not suppressed by the mechanical response, but in the resolved sideband regime, i.e.~$\omega_m\gg\kappa$, this process is highly off-resonant with respect to the cavity frequency. Thus, the effect at this frequency is small. As shown in the insets of Fig.~\ref{pltOMIT}, dissipative coupling leads to a larger contribution at this detuning $\delta\approx-\omega_m$ than purely dispersive coupling. This originates from the direct interaction between optical bath and mechanical oscillator: It gives rise to a constant contribution to $\alpha(\delta)^2$ in Eq.~(\ref{eqA-}), i.e.~a term not filtered by the cavity response function.

In contrast, if $\delta=+\omega_m$ and $\Delta=-\omega_m$, the probe frequency $\omega_p$ coincides with the cavity resonance $\omega_c$, giving rise to more prominent effects. The optomechanical coupling leads to a narrow peak enclosed by a broad dip that appears also in absence of coupling. For small coupling as shown in Figs.~\ref{pltOMIT} (a) and (c), the width of this peak is given by the width of the mechanical resonance. The mechanical linewidth, in turn, is given by the intrinsic damping $\gamma$ and broadened with increasing coupling strength due to the additional optical damping $\gamma_\mathrm{opt}$. In the case of sufficiently strong coupling, see Figs.~\ref{pltOMIT} (b) and (d), the two modes are separated by a peak that has a width comparable to or larger than the width of each of the modes. The splitting increases for stronger coupling.

This general behaviour is shared by dissipatively and dispersively coupled systems, but there are small differences: First, the width of the splitting in case of purely dissipative and purely dispersive coupling depends differently on the respective coupling strength. Second, there is an increasing asymmetry between the two modes in case of dissipative coupling, whereas purely dispersive coupling leads to a splitting into two anti-peaks that remain similar over a larger range of coupling strengths.

In analogy to the treatment of purely dispersive coupling \cite{wrdgask2010}, we assume that only anti-Stokes scattering occurs. This can be described by simplified equations of motion where we neglect coupling to $\hat{d}^\dagger$ and $\hat{c}^\dagger$ in Eqs.~(\ref{eqLinEOMc}) and (\ref{eqLinEOMd}). As a result $\langle\hat{\xi}_\mathrm{out}(t)\rangle$ is still of the form of Eq.~(\ref{eqXiOut}), but with new coefficients $A_\mathrm{approx}^+ =0$ and
\begin{equation}\label{eqAapprox}
A_\mathrm{approx}^- =\left[1-\kappa\chi_c(\delta)-\kappa\frac{|\bar{a}|^2\alpha(\delta)^2}{\chi_m^{-1}(\delta)+i\tilde{\Sigma}(\delta)}\right]\bar{d}_\mathrm{probe}
\end{equation}
where $\tilde{\Sigma}(\omega)=\tilde{\Sigma}_{\tilde{A}}(\omega)+\tilde{\Sigma}_{\tilde{B}}(\omega)+\tilde{\Sigma}_{\tilde{A}\tilde{B}}(\omega)$. Here, $\tilde{\Sigma}_{\tilde{A}}(\omega) =-i(\tilde{A}\kappa|\bar{a}|)^2\chi_c(\omega)$, $\tilde{\Sigma}_{\tilde{B}}(\omega) = i(\tilde{B}/2)^2\left|\bar{a}\right|^2 \chi_c(\omega)(i\Delta+\kappa/2)^2$, and $\tilde{\Sigma}_{\tilde{A}\tilde{B}}(\omega) =\tilde{B} \tilde{A}\kappa\left|\bar{a}\right|^2\chi_c(\omega)(i\Delta+\kappa/2)$ denote only the parts of the originally defined self-energies with weight at $\delta\approx+\omega_m$. In the case of $\tilde{B}=0$ the approximation is the same as in Ref.~\cite{wrdgask2010}. For both dispersive and dissipative coupling, the approximation is valid in the resolved sideband regime, e.g.~it does not reproduce the feature at $\delta=-\omega_m$ which becomes more important if $\omega_m$ is of the order of $\kappa$ or the coupling becomes too strong.

\section{Conclusion} \label{sec:conclusion}

We have presented a detailed study of optomechanical systems featuring both dissipative and dispersive coupling. For weak coupling we have employed a quantum noise approach to calculate the optical damping and the optically-induced frequency shift. Surprisingly, there are two regions leading to cooling and two regions leading to amplification. This is a consequence of the Fano line shape in the force spectrum which is absent for purely dispersive coupling. In the strong-coupling regime we have calculated the mechanical and the optical spectra from the exact solution to the linearized equations of motion. Similar to purely dispersive coupling, normal-mode splitting appears for sufficiently strong coupling. Nonzero dissipative coupling additionally leads to a striking feature which originates from quantum noise interference. Finally, we have found purely dissipative coupling can lead to optomechanically-induced transparency which will be an experimentally convenient way to observe normal-mode splitting.

\section*{Acknowledgements}

We thank A.~A.~Clerk, S.~M.~Girvin, K.~Hammerer, T.~J.~Kippenberg, M.~Ludwig, F.~Marquardt, and P.~Treutlein for interesting discussions. This work was financially supported by the Swiss NSF and the NCCR Quantum Science and Technology.

\end{document}